\documentclass[12pt]{article}

\usepackage{soul}

\usepackage{tikz}
\usepackage{epsfig,latexsym,amsfonts,amsmath,amsthm,amssymb,amsbsy,multirow,slashed,wasysym,textcomp,subfigure,wrapfig,datetime,comment,mathtools,cancel,cite,twistor,mathrsfs}
\usepackage[hidelinks]{hyperref}
\usepackage[font={footnotesize},bf]{caption}

\def\ee{\end{equation}}
\def\be{\begin{equation}}
\def\bea{\begin{eqnarray}}
\def\eea{\end{eqnarray}}
\newcommand{\beq}{\begin{eqnarray}}
\newcommand{\eqq}{\end{eqnarray}}
 \newcommand{\badat}{\begin{alignedat}}
 \newcommand{\eadat}{\end{alignedat}}
\newcommand{\bh}{\bar{h}}
\newcommand{\eal}[1]{\be \begin{aligned} #1 \end{aligned}\end{equation}} 
\newcommand{\eqn}[1]{\be #1 \end{equation}} 
\newcommand{\eqa}[1]{\bea  #1\end{eqnarray}}
\newcommand{\CT}{\mathcal{CT}^2}
\renewcommand{\d}{\mathrm{d}}

\long\def\new#1\endnew{{\bf #1}}		
\long\def\del#1\enddel{}
\def\eps{\epsilon }
\def\del{\partial}

\def\re{\mathrm{e}}

\usepackage{color}

\newcommand{\pink}[1]{\textcolor{\pink}{#1}}

\definecolor{dblue}{rgb}{0.2,0.50,0.80}

\newcommand{\zb}{\bar{z}}

\def\O{\mathcal{O}}

\def\D{{\Delta}}

\def\G{{\Gamma}}

\def\vphi{\varphi}

\def\AdS{\text{AdS}}
\newcommand{\br}[1]{\overline{#1}}
\newcommand{\hb}{\bar{h}}

\def\cL{{\cal L}}
\def\bh{{\bar h}}
\def\bz{{\bar z}}
\def\bw{{\bar w}}

\def\s{ {\sigma} }

\newcommand{\bs}{\bar{\sigma}}

 \def\e{\epsilon}

\newcommand{\MHV}{\mathrm{MHV}}

\numberwithin{equation}{section} 

\begin{document}
\begin{titlepage}
  \thispagestyle{empty}
  \begin{flushright}
    \end{flushright}
  \bigskip
  \begin{center}
	 \vskip2cm
  \baselineskip=13pt {\LARGE \scshape{
  \vspace{0.5em} Soft Algebras for Leaf Amplitudes }}

	 \vskip2cm
   \centerline{Walker Melton, Atul Sharma and Andrew Strominger}
 \vskip.5cm
 \noindent{\em Center for the Fundamental Laws of Nature,}
  \vskip.1cm
\noindent{\em  Harvard University,}
{\em Cambridge, MA, USA}
\bigskip
  \vskip1cm
  \end{center}
  \begin{abstract}
Celestial MHV amplitudes are comprised of non-distributional leaf amplitudes associated to an AdS$_3$ leaf of a foliation of flat spacetime.  It is shown here that the leaf amplitudes are governed by the same infinite-dimensional soft `$S$-algebra' as their celestial counterparts. Moreover, taking the soft limit of the smooth three-point MHV leaf amplitude yields a nondegenerate minus-minus two-point leaf amplitude. The two- and three-point MHV leaf amplitudes are used to compute  the  plus-minus-minus leaf operator product coefficients. 
   \end{abstract}
%

%
\end{titlepage}
\tableofcontents

\section{Introduction} 
Leaf amplitudes \cite{Melton:2023bjw} are simple building blocks of celestial amplitudes in which the particle interaction vertex is integrated over only one leaf of a hyperbolic foliation of spacetime. As such they share the 2D conformal invariance, but not the translation invariance, of celestial amplitudes. The latter enforces a distributional structure on the celestial amplitudes, while the leaf amplitudes from which they are composed are given by generally smoother and well-understood AdS-Witten diagrams \cite{Law:2019glh, Melton:2023bjw}.
For the special case of MHV tree leaf amplitudes, the celestial amplitudes have a remarkably simple expression as the residue of a pole in the leaf amplitudes\cite{Melton:2023bjw}.\footnote{The formula for the most general celestial amplitude in terms of leaf amplitudes has not been worked out, but presumably involves more involved compositions of products of leaf amplitudes. Relevant recent progress appears in \cite{Hao:2023wln,Iacobacci:2024nhw}.}

In addition to translation invariance, celestial amplitudes are constrained by  $w$-symmetries (for gravity) and $S$-symmetries (for gauge theory) arising from infinite towers of soft theorems \cite{Guevara:2021abz,Strominger:2021mtt}. While leaf amplitudes are certainly not translation invariant,  it is not a priori  obvious if they realize these symmetries. 
The main result of this paper is to show that MHV tree amplitudes indeed transform under the $S$-algebra generated by soft insertions. 

Several interesting relations are derived along the way to this result.  Simple expressions are given for all MHV leaf amplitudes in terms of AdS-Witten contact diagrams and used to derive the leading soft theorem. Specializing to three points, we derive a simple expression for the minus-minus leaf  two-point function which, in sharp contrast to its celestial analog, is nonvanishing.  Hence  leaf amplitudes  more closely resemble correlators  of  a standard 2D CFT. This enables the application of standard  formulae  to derive the complete set of plus-minus-minus single-gluon leaf operator product coefficients, which we show agrees with a collinear analysis at higher points.

Throughout this paper we work in $(2, 2)$ signature  Klein space. This importantly enables `holomorphic' expansions in which the celestial coordinate $\bz$ remains fixed while $z\to 0$ \cite{Atanasov:2021oyu}.  Lorentzian scattering amplitudes may be  obtained by analytic continuation at the end of the Kleinian computation.

A number of relevant recent works study  both the decomposition of scattering amplitudes as AdS-Witten diagrams \cite{Pasterski:2016qvg,Casali:2022fro,Iacobacci:2022yjo} and 
the construction of a holographic dictionary lifting in various ways the analogy from AdS \cite{deBoer:2003vf,deGioia:2022fcn,Hao:2023wln,Sleight:2023ojm,Jain:2023fxc, Bagchi:2023fbj,Iacobacci:2024nhw}. These efforts commonly  seek to apply our deep and systematic  understanding of AdS holography to flat space holography.

This paper is organized as follows. Preliminaries and conventions, including the geometry of Klein space and the definition of leaf amplitudes, are  in section \ref{sec:pre}. Section \ref{sec:leading} derives the leading soft gluon theorem for MHV leaf trees from their representation as AdS-Witten diagrams. In section \ref{sec:2pt} we derive the two-point minus-minus leaf amplitude and find that it takes the familiar form of a CFT two-point function that is single valued on the celestial torus. In section \ref{sec:ope} the general plus-minus-minus leaf OPE is derived using the three- and two-point MHV leaf amplitudes.  This result is compared and found to nontrivially agree with the leading term in a collinear analysis of higher-point leaf amplitudes after using a contour deformation to evaluate it as a sum over poles. Finally in section \ref{sec:S} the commutators of the leaf soft currents are shown to obey the same $S$-algebra as their celestial counterparts. 


\section{Preliminaries}
\label{sec:pre}

In this section, we recall our construction of leaf amplitudes that was introduced in \cite{Melton:2023bjw}. This is most naturally done in Klein space $\R^{2,2}$, i.e., $(2,2)$ signature flat space. 

\subsection{Celestial amplitudes in Klein space}
\label{ssec:celamp}

An introduction to the Kleinian geometry relevant for our purposes may be found in \cite{Atanasov:2021oyu,Melton:2023hiq,Melton:2023bjw}. We will denote Cartesian coordinates on $\R^{2,2}$ by $X^\mu$, and take the metric of $\R^{2,2}$ to be the flat metric of $(--+\,+)$ signature. In this signature, null infinity has a single connected component and is foliated by celestial tori. Each celestial torus $\CT=S^1\times S^1$ may be viewed (non-canonically) as a union of two Lorentzian diamonds $\R^{1,1}$ equipped with light cone coordinates $z,\bz$ that are real and independent. The Lorentz group of Klein space $\SL(2,\R)\times\br{\SL}(2,\R)$ acts by real and independent M\"obius transformations on $z,\bz$. The latter constitute the conformal group of $\CT$ \cite{Atanasov:2021oyu}.
To study CFT correlators on such a Lorentzian torus, it turns out to be much easier to work with global coordinates $\s,\bs$ defined by setting
\be
z = \tan\s\,,\qquad\bz = \tan\bs\,.
\ee
They are periodically identified under $(\s,\bs)\sim(\s+2\pi,\bs)$ and $(\s,\bs)\sim(\s+\pi,\bs+\pi)$. The torus is covered exactly once by the fundamental domain $\sigma\in[0,2\pi)$, $\bs\in[0,\pi)$.

Null momenta $p_i^\mu$ in $(2,2)$ signature can be parametrized by frequencies $\omega_i>0$ and points $(\s_i,\bs_i)\in\CT$ through the following relations:
\be
\begin{split}
    p_i^\mu &= \omega_i\hat p_i^\mu\,,\\
    \hat p_i^\mu &= (\cos\psi_i,\sin\psi_i,\cos\vphi_i,\sin\vphi_i)\,,\\
    \psi_i &\vcentcolon= \s_i+\bs_i\,,\qquad\vphi_i\vcentcolon=\s_i-\bs_i\,.
\end{split}
\ee
In what follows, we will use the abbreviations
\be
s_{ij} \vcentcolon= \sin\s_{ij}\,,\qquad\bar s_{ij}\vcentcolon=\sin\bs_{ij}\,.
\ee
They are the natural variables that enter 2D conformal correlators on the torus. For instance, we find the Lorentz invariants $\hat p_i\cdot\hat p_j = 2\sin\s_{ij}\sin\bs_{ij}$.

If $A(1^{J_1}2^{J_2}\cdots n^{J_n})$ denotes the scattering amplitude of momentum eigenstates carrying null momenta $p_i$ and helicities $J_i$, the corresponding celestial amplitude is defined by taking its Mellin transform,
\be
\cA(1^{J_1}2^{J_2}\cdots n^{J_n}) = \prod_{i=1}^n\int_0^\infty\d\omega_i\,\omega_i^{\Delta_i-1}\,A(1^{J_1}2^{J_2}\cdots n^{J_n})\,.
\ee
It transforms as a CFT correlator of operators carrying conformal weights
\be
(h_i,\bh_i) = \left(\frac{\Delta_i+J_i}{2},\frac{\Delta_i-J_i}{2}\right)
\ee
in a 2D CFT living on $\CT$, the so-called celestial CFT (CCFT). Defining celestial amplitudes using global coordinates as we have done here makes manifest their single-valuedness on the celestial torus. It also removes the need for the ``incoming/outgoing'' labels that the reader may have encountered in past work on celestial holography.


\subsection{Leaf amplitudes}
\label{ssec:leaf}

We will be interested in studying a refinement of celestial amplitudes that we call \emph{leaf amplitudes}. These are motivated by the hyperbolic foliation of Klein space. 

$\R^{2,2}$ can be decomposed into a timelike wedge on which $X^2<0$ and a spacelike wedge on which $X^2>0$. On each wedge, we can write $X^\mu=\tau\hat x^\mu$, where $\tau>0$ and $\hat x^2=-1$ or $+1$. In either case, slices of constant $\tau$ are copies of Lorentzian $\AdS_3/\Z$. Taken together, they foliate Klein space.

In this work, we will be interested in the leaf amplitudes that build up tree-level MHV gluon amplitudes. These are defined as follows. The color-stripped MHV gluon amplitude is given by the Parke-Taylor formula \cite{PhysRevLett.56.2459}
\be
\begin{split}
    A(1^-2^-3^+\cdots n^+) &= \frac{\la12\ra^3}{\la23\ra\la34\ra\cdots\la n1\ra}\;\delta^4\bigg(\sum_{i=1}^n p_i\bigg)\\
    &= \frac{\la12\ra^3}{\la23\ra\la34\ra\cdots\la n1\ra}\int_{\R^{2,2}}\frac{\d^4X}{(2\pi)^4}\;\re^{\im\sum_jp_j\cdot X}\,,
\end{split}
\ee
which is expressed in terms of the spinor-helicity brackets
\be
\la ij\ra \vcentcolon= \sqrt{\omega_i\omega_j}\,s_{ij}\,.
\ee
To get the second line, we have substituted the Fourier representation of the momentum conserving delta function. Breaking this into integrals over the timelike and spacelike wedges, we can integrate out $\tau$ -- the magnitude of $X^\mu$ -- and perform the Mellin transforms to find
\be
\cA(1^-2^-3^+\cdots n^+) = \frac{\delta(\beta)}{8\pi^3}\,\big(\cL(\s_i,\bs_i) + \cL(\s_i,-\bs_i)\big)\,,
\ee
where $\beta=\sum_i(\Delta_i-1)$. This expands the celestial amplitude in terms of constituents defined along the leaves of the hyperbolic foliation,
\be\label{leaf}
\cL(\s_i,\bs_i) = \frac{s_{12}^3}{s_{23}s_{34}\cdots s_{n1}}\int_{\hat x^2=-1}{\rm D}^3\hat x\;\prod_{i=1}^n\frac{\Gamma(2\bh_i)}{(-\im\hat p_i\cdot\hat x+\eps)^{2\bh_i}}\,,
\ee
where $\eps>0$ is a small regulator needed to make the Mellin integrals converge, and ${\rm D}^3\hat x$ is a measure on $\AdS_3/\Z$. We refer to $\cL(\s_i,\bs_i)\equiv\cL(1^-2^-3^+\cdots n^+)$ as \emph{leaf amplitudes}. 

Note that in defining $\cL$, we have chosen to continue the MHV prefactor $\langle ij\rangle^4/\langle 12\rangle\cdots\langle n1 \rangle$ off the support of momentum conservation in a particular way; this continuation is not unique. Different leaf amplitudes can give the same celestial amplitude: for example the addition of total derivatives to the action  may affect the former but not the latter. It is possible however that imposing soft relations will significantly constrain expressions for $\cL$.

Leaf amplitudes are conformally covariant and in the MHV case considered herein given by contact Witten diagrams of $\AdS_3/\Z$. We will formally write them as correlation functions of spin 1 conformal primaries $\O^-_{\Delta_i}(\s_i,\bs_i)$ whenever convenient:
\begin{equation}
    \cL(1^-2^-3^+\cdots n^+) = \langle \O^-_{\Delta_1}(\s_1,\bs_1)\, \O^-_{\Delta_2}(\s_2,\bs_2) \cdots \O^+_{\Delta_n}(\s_n,\bs_n)\rangle_{\cL}\,.
\end{equation}
We expect that leaf amplitudes will prove to be a useful ingredient in looking for celestial duals. One important property of leaf amplitudes is that they remain non-distributional at 3-points, a trait that is not shared by the full amplitude. This is because leaf amplitudes are not required to satisfy the usual constraints of 4D translation invariance. This was leveraged in \cite{Melton:2023bjw} to find a decomposition of the 3-gluon celestial amplitude in terms of non-distributional building blocks.


\section{Leading soft theorem}
\label{sec:leading}

In this section we give a simple derivation of the leading soft theorem for leaf amplitudes. 

Gauge theories obey universal factorization rules that govern the simple  pole in $\omega$ in gluon scattering amplitudes. Once transformed to the conformal basis, the leading soft theorem governs the residue of celestial amplitudes as $\Delta \to 1$ \cite{Donnay:2018neh,Pate:2019mfs}. In this section, we show that  $n$-gluon MHV leaf amplitudes obey the same leading soft theorem as full celestial amplitudes. In the MHV sector, only positive particles  can nontrivially be taken  soft.

We can make particle $n$ soft in the MHV leaf amplitudes defined in \eqref{leaf} by taking a residue at $\Delta_n=1$. Since it is a positive helicity gluon, this is equivalent to taking a residue at $\bh_n = (\Delta_n-1)/2=0$. The soft limit simply extracts the residue of the $\Gamma(2\bh_n)$ factor present in the integrand, and the wavefunction $(\eps-\im\hat p_n\cdot\hat x)^{-2\bh_n}$ drops out. One finds
\begin{equation}
\begin{split}
\lim_{\Delta_n\to 1} (\Delta_n-1)\,\mathcal{L}(1^-2^-3^+\cdots n^+) &= \frac{s_{12}^3}{s_{23}\cdots s_{n1}}\lim_{\bh_n\to 0} 2\bh_n\int {\rm D}^3\hat{x}\prod_{j=1}^n\frac{\Gamma(2\bar{h}_j)}{(\epsilon-\im\hat p_j\cdot\hat{x})^{2\bar{h}_j}} \\
&= \frac{s_{12}^3}{s_{23}\cdots s_{n1}}\int {\rm D}^3\hat{x}\prod_{j=1}^{n-1}\frac{\Gamma(2\bar{h}_j)}{(\epsilon-\im\hat p_j\cdot\hat{x})^{2\bar{h}_j}} \\
&= \frac{s_{n-1,1}}{s_{n-1,n}s_{n1}}\, \mathcal{L}(1^-2^-3^+\cdots (n-1)^+)
\end{split}
\end{equation}
replicating the leading conformally soft theorem for $n$-gluon color-stripped MHV amplitudes.
The full soft theorem may then be reproduced by summing over color orderings.

We expect that a similar derivation of the higher subleading soft relations along the same lines should be possible. One encounters positive powers of $(\eps-\im\hat p_n\cdot\hat x)$ which may be transformed into derivatives acting on the $(n-1)$ point amplitude. However we instead give below 
a less tedious and more illuminating derivation of these relations in a collinear limit after deriving  the OPEs.


\section{Two-point leaf amplitudes}
\label{sec:2pt}

This section considers the especially interesting case of soft factorization of the 3-point leaf amplitude. It gives an important new clue about the structure of celestial CFT: a non-vanishing 2-point leaf amplitude. In the MHV sector, one obtains a 2-gluon minus-minus leaf amplitude $\cL(1^-2^-)$ which has the structure of a generically smooth 2-point CFT correlator. In the following section we exploit this to construct the OPEs from the 3-point leaf amplitude.  At three points, we can work with the explicit expression for the 3-gluon leaf amplitude \cite{Melton:2023bjw}
\begin{align}\label{L3}
    \mathcal{L}(1^-2^-3^+) &= \frac{\im\pi\cN}{2}\frac{s_{12}^3}{s_{23}s_{31}}\left[\frac{1}{(s_{12}\bar{s}_{12}+\im\e)^{\bh_1+\bh_2-\bh_3}(s_{23}\bar{s}_{23}+\im\e)^{\bh_2+\bh_3-\bh_1}(s_{31}\bar{s}_{31}+\im\e)^{\bar{h}_3+\bar{h}_1-\bar{h}_2}} \right.\nonumber\\
    &\hspace{2.4cm}-\left.\frac{1}{(s_{12}\bar{s}_{12}-\im\e)^{\bh_1+\bh_2-\bh_3}(s_{23}\bar{s}_{23}-\im\e)^{\bh_2+\bh_3-\bh_1}(s_{31}\bar{s}_{31}-\im\e)^{\bar{h}_3+\bar{h}_1-\bar{h}_2}}\right]\nonumber \\[0.5em]
    \cN &= \Gamma\bigg(1+\frac\beta2\bigg)\,\Gamma(\hb_1+\hb_2-\hb_3)\,\Gamma(\hb_{12}+\hb_3)\,\Gamma(\hb_{21}+\hb_3)
\end{align}
where \be \beta = \Delta_1+\Delta_2+\D-3 = 2\,(\bh_1+\bh_2+\bh_3-2) \ee and $\bh_{ij}\vcentcolon=\bh_i-\bh_j$ and we use the branch cut prescription given in \cite{Melton:2023hiq}. This implies  that the two terms in 
\eqref{L3} do $not$ generically cancel because of differing phases acquired when crossing various branch cuts where $s_{ij}$ vanishes. 

To compute the soft limit of  \eqref{L3} we use  the delta function identity
\begin{equation}
    \lim_{a\to 0}\frac{\Gamma(a+\im x)\,\Gamma(a-\im x)}{\Gamma(2a)} = 2\pi\,\delta(x)\,,
\end{equation}
which, for weights on the principal series $\hb_j = 1+\im\lambda_j/2$, yields
\begin{equation}
    \lim_{\bar{h}_3\to 0}\frac{\Gamma(\hb_{12}+\hb_3)\,\Gamma(-\hb_{12}+\hb_3)}{\Gamma(2\hb_3)} = 2\pi\,\delta(\im\hb_{12})\,.
\end{equation}
Equivalently
\begin{equation}
    \begin{split}
        \lim_{\hb_3\to 0} \frac{\cN}{\Gamma(2\hb_3)} = 2\pi\, \delta(\im\hb_{12})\,\Gamma(2\hb_1-1)\,\Gamma(2\hb_1)
    \end{split}
\end{equation} Using this relation we find the soft limit
\begin{align}
    &\lim_{\Delta_3\to1}(\Delta_3-1)\,\cL(1^-2^-3^+) = \lim_{\hb_3\to 0}\frac{\cL(1^-2^-3^+)}{\Gamma(2\hb_3)} \nonumber\\
    &= \frac{s_{12}}{s_{23}s_{31}}\left[\im\pi^2 \delta(\im\hb_{12})\,\Gamma(2\hb_1-1)\,\Gamma(2\hb_1)\left(\frac{s_{12}^2}{(s_{12}\bar{s}_{12}+\im\e)^{2\hb_1}}- \frac{s_{12}^2}{(s_{12}\bar{s}_{12}-\im\e)^{2\hb_1}}\right)\right]\,.
\end{align}
Remarkably, this has the structure of a soft-prefactor $s_{12}/s_{23}s_{31}$ times a {standard} 2-point CFT correlator of weight $\Delta_1=h_1+\hb_1$ and spin $-1$. 

From this, we can read off the minus-minus two-point leaf amplitude, recalling that negative  helicity gluons have $\bar{h} -h= 1$,
\begin{equation}\label{L2}
    \begin{split}
        \mathcal{L}(1^-2^-) &= \langle \O^-_{\Delta_1}(\s_1,\bs_1)\,\O^-_{\Delta_2}(\s_2,\bs_2)\rangle_{\cL}\\
        &= 2\pi^2\im\,\delta(\Delta_1-\Delta_2)\,\Gamma(\Delta_1)\,\Gamma(\Delta_1+1)\left[\frac{s_{12}^2}{(s_{12}\bar{s}_{12}+\im\e)^{\Delta_1+1}}- \frac{s_{12}^2}{(s_{12}\bar{s}_{12}-\im\e)^{\Delta_1+1}}\right]
    \end{split}
\end{equation}
where we have suppressed the $\im$'s in $\delta(\Delta_1-\Delta_2)$ for notational ease. This result is consistent with conformal covariance for $\epsilon \to 0$. In particular, it is non-zero precisely when $\Delta_1=\Delta_2$ as expected from conformal Ward identities. 
As a check on our result, it can also be obtained directly from the definition \eqref{leaf} for $n=2$,
\begin{equation}
    \cL(1^-2^-) = -s_{12}^2\int_{\hat x^2=-1}{\rm D}^3\hat x\;\prod_{i=1}^2\frac{\Gamma(2\bh_i)}{(-\im\hat p_i\cdot\hat x+\eps)^{2\bh_i}}\,.
\end{equation}

As a consistency check, from this leaf amplitude we can compute the minus-minus celestial amplitude:
\be
    \cA(1^-2^-) = \frac{\delta(\beta)}{8\pi^3}\,\big(\cL(\s_i,\bs_i) + \cL(\s_i,-\bs_i)\big)\,.\\
\ee The only common support of $\delta(\beta)$ and $\delta(\Delta_1-\Delta_2)$ is $\Delta_1=\Delta_2=1$. For these weights, the factor in the brackets in \eqref{L2} becomes
\be
\frac{s_{12}^2}{(s_{12}\bar{s}_{12}+\im\e)^{2}}- \frac{s_{12}^2}{(s_{12}\bar{s}_{12}-\im\e)^{2}} = -2\pi\im\,\sgn(s_{12})\,\delta'(\bar s_{12})\,.
\ee
As a result, the minus-minus celestial amplitude vanishes as expected
\be
\cA(1^-2^-) = \frac{1}{4}\,\delta(\Delta_1-1)\,\delta(\Delta_2-1)\,\sgn(s_{12})\big[\delta'(\bar s_{12})+\delta'(-\bar s_{12})\big] = 0\,.
\ee

This is an elementary example of a generic phenomenon.  Leaf correlators  are in general nonvanishing in accord with garden-variety CFT expectations. However the construction of celestial from leaf correlators often involves cancellations  required by translation invariance \cite{Law:2019glh}.


\section{OPE for leaf amplitudes}
\label{sec:ope}

In this section, we investigate the OPE structure of leaf amplitudes from two perspectives: the chiral collinear structure of $n$-point MHV leaf amplitudes and  the relationship between the 2- and 3-point leaf amplitudes. While the two perspectives at first seem to give different results, we show that they are compatible to leading order by performing a contour deformation. We work in planar  coordinates throughout this section. 


\subsection{OPEs from Three-Point Structures}
\label{ssec:ope3pt}

In a general CFT, the 3-point function is intimately connected to the operator product expansion. In particular, if three primary operators $\O_i$ have 3-point function
\begin{equation}
    \big\langle \O_1(z_1,\zb_1)\,\O_2(z_2,\zb_2)\,\O_3(z_3,\zb_3)\big\rangle = \frac{C_{123}}{z_{12}^{h_1+h_2-h_3}\zb_{12}^{\hb_1+\hb_2-\hb_3}\times\mathrm{cyclic}}
\end{equation}
and 2-point function
\begin{equation}
    \big\langle \O_j(z_1,\zb_1)\,\O_j(z_2,\zb_2) \big\rangle = \frac{D_{jj}}{z_{12}^{2h}\zb_{12}^{2\hb}}\,,
\end{equation}
then the operator product expansion will contain a term of the form \cite{DiFrancesco:639405}
\begin{equation}
    \O_1(z_1,\zb_1)\,\O_2(z_2,\zb_2) \sim \frac{C_{12}^{\ \ 3}}{z_{12}^{h_1+h_2-h_3}\zb_{12}^{\hb_1+\hb_2-\hb_3}}\,\O_3(z_2,\zb_2)\,,\qquad C_{12}^{\ \ 3} = \frac{C_{123}}{D_{33}}\,.
\end{equation}
The exact OPE has corrections involving $\p$ and $\bar \p$ derivatives of primaries determined by $\SL(2,\R)\times\br{\SL}(2,\R)$ conformal symmetry. 
Generalizing this to the case of a continuous spectrum for leaf amplitudes gives the OPE \begin{equation}
\label{eq:contope}
    \O^+_{\Delta_1}\O^-_{\Delta_2} \sim -\im\int\d\Delta \;z^{\frac{\Delta-\Delta_1-\Delta_2-1}{2}}\zb^{\frac{\Delta-\Delta_1-\Delta_2+1}{2}} \,C^{+--,\Delta}_{\Delta_1,\Delta_2}\O^-_{\Delta}(z_2,\zb_2) + \cdots
\end{equation}
where  $\cdots$ includes terms other than negative-helicity single-gluon operators and $\p, \bar \p$ descendants of the primary operators. Plugging this ansatz into the 3-point function \eqref{L3} and making use of the 2-point function \eqref{L2} shows that
\begin{equation}\label{C+--}
    C^{+--,\Delta}_{\Delta_1,\Delta_2} = \frac{1}{4\pi}\frac{\Gamma\left(\frac{\Delta_1+\Delta_2+\Delta-1}{2}\right)\Gamma\left(\frac{\Delta_1+\Delta_2-\Delta-1}{2}\right)\Gamma\left(\frac{\Delta_1-\Delta_2+\Delta-1}{2}\right)\Gamma\left(\frac{\Delta_2-\Delta_1+\Delta+3}{2}\right)}{\Gamma(\Delta)\Gamma(\Delta+1)}.
\end{equation}
 In this integral, $\Delta$ is integrated along the continuous principal series $\Delta = 1 +\im\lambda$ and the integration variable is $\d \lambda = -\im\, \d\Delta$.
 At the poles of $C^{+--\Delta}_{\Delta_1,\Delta_2}$, it is chosen to pass to the left of $\Delta = \Delta_1+\Delta_2-1$ and to the right of $\Delta = \Delta_2-\Delta_1+1$. 

In the $\Delta$ plane, the OPE coefficient \eqref{C+--} has poles at 
\begin{align}
    \Delta &= \Delta_1 +\Delta_2 - 1+ 2\mathbb{Z}_{\ge 0},\label{asc} \\
    \Delta &= \Delta_2-\Delta_1 + 1 - 2\mathbb{Z}_{\ge 0}, \\
    \Delta &= \Delta_1-\Delta_2-3 - 2\mathbb{Z}_{\ge 0} ,\\
    \Delta &= -\Delta_1-\Delta_2+1-2\mathbb{Z}_{\ge 0}.
\end{align}
Deforming the $\Delta$ contour to surround the ascending series of poles \eqref{asc} implies that
\begin{align}
    &\O^+_{\Delta_1}(z_1,\zb_1)\,\O^-_{\Delta_2}(z_2,\zb_2) \sim -2\pi \sum_{n=0}^\infty z_{12}^{n-1}\zb_{12}^n\,\mathrm{Res}_{\Delta = \Delta_1 + \Delta_2 - 1 + 2n} C^{+--,\Delta}_{\Delta_1,\Delta_2}\O^-_\Delta(z_2,\zb_2)\nonumber \\
    &\sim \frac{1}{z_{12}}B(\Delta_1-1,\Delta_2+1)\,\O^-_{\Delta_1+\Delta_2-1}(z_2,\zb_2)\label{defope}\\
    &\quad+ \sum_{n=1}^\infty z_{12}^{n-1}\zb_{12}^n \,\frac{(-1)^nB(\Delta_1+n-1,\Delta_2+n+1)\Gamma(\Delta_1+\Delta_2+n-1)}{\G(n+1)\Gamma(\Delta_1+\Delta_2+2n-1)}\; \O^-_{\Delta_1+\Delta_2+2n-1}(z_2,\zb_2)+ \cdots\nonumber  
\end{align}
where $\cdots$ includes non-singular terms that may include multi-particle operators as well as descendants.  Note that the contour in $\Delta$ must be chosen so that the ascending series of poles ($\Delta = \Delta_1+\Delta_2-1+2n$) is on the right while the descending series of poles are on the left. The collinear $z_{12}$ pole term displayed in the second line of \eqref{defope} matches that of the celestial OPE derived from the full MHV amplitude \cite{Stieberger:2018onx,Pate:2019lpp}. 
 The nonsingular terms in the last two lines resemble celestial OPE corrections from higher-dimension corrections to the bulk action \cite{Pate:2019lpp}, but 
in this context are not generated by a translation-invariant correction. 

\subsection{Leading OPE from Collinear Singularities}
\label{ssec:collinear}

In this section we show that the leading $z_{12}$ pole can be directly found from the singularity in the collinear expansion of higher-point MHV leaf amplitudes. 
In $z,\bz$ coordinates\footnote{Because the affine $z,\bz$ coordinates cover half of the celestial torus, leaf amplitudes written in these coordinates should possess a label describing which half the operator is inserted in. Here, we assume that the operators that are becoming collinear are in the same coordinate patch and suppress this label}, we have that
\begin{equation}
\begin{split}
    \mathcal{L}(1^-2^-\cdots n^+) &= \MHV(1^-2^-3^+\cdots n^+) \int{\rm D}^3\hat{x}\prod_{i=1}^n\frac{\Gamma(2\hb_i)}{(\eps-\im \hat{p}_j \cdot \hat{x})^{2\hb_i}} \\
    \MHV(1^-2^-\cdots n^+) &= \frac{z_{12}^3}{z_{23}\cdots z_{n1}}
\end{split}
\end{equation}
We examine both the collinear limit $z_{34} \to 0$, which will give us the $++ \to +$ OPE, and the $z_{23} \to 0$ limit, which will give us the $-\,+ \to -$ OPE.  In each case, we use the chiral collinear expansion for the product of conformal primary wavefunctions given by\footnote{It would be interesting to try to reproduce the full expression \eqref{defope} from the subleading terms in this collinear expansion.} \cite{Casali:2022fro} 
\begin{equation}
\label{eq:primwavexp}
    \begin{split}
        \frac{\Gamma(2\hb_1)}{(\eps-\im\hat{p}_1\cdot\hat{x})^{2\hb_1}}\frac{\Gamma(2\hb_2)}{(\eps-\im \hat{p}_2\cdot\hat{x})^{2\hb_2}} = \sum_{j=0}^\infty \frac{\Gamma(2\hb_1+j)\,\Gamma(2\hb_2)}{j!\,\Gamma(2\hb_1+2\hb_2+j)}\,\zb_{12}^j \bar{\partial}_2^j \frac{\Gamma(2(\hb_1+\hb_2))}{(\eps-\im\hat{p}_2\cdot\hat{x})^{2(\hb_1+\hb_2)}} + O(z_{12})
    \end{split}
\end{equation}


\subsubsection{$++ \to +$}
\label{sssec:+++}

Taking the limit $z_3 \to z_4$, we have that 
\begin{equation}
    \MHV(1^-2^-3^+4^+\cdots n^+) \to \frac{1}{z_{34}}\,\MHV(1^-2^-4^+\cdots n^+)
\end{equation}
Using the expansion \eqref{eq:primwavexp}, we find
\begin{equation}
\label{eq:ppope}
    \O^+_{\Delta_3}(z_3,\zb_3)\,\O^+_{\Delta_4}(z_4,\zb_4) \sim \frac{1}{z_{34}}\sum_{j=0}^\infty\frac{\Gamma(\Delta_3+j-1)\Gamma(\Delta_4-1)}{\Gamma(\Delta_3+\Delta_4+j-2)j!}\,\zb_{34}^j\bar{\partial}_4^j \O^+_{\Delta_3+\Delta_4-1}(z_4,\zb_4) + O(z_{34}^0)
\end{equation}
when inserted in an $n$-point MHV leaf amplitude.


\subsubsection{$-+\to-$}
\label{sssec:-+-}

In the limit $z_2 \to z_3$, we similarly have
\begin{equation}
    \MHV(1^-2^-3^+\cdots ) = \frac{1}{z_{23}}\, \MHV(1^-2^-4^+\cdots)
\end{equation}
Using \eqref{eq:primwavexp}, this implies that, inside an $n$-point MHV leaf amplitude,  
\begin{equation}
\label{eq:pmope}
    \O^-_{\Delta_2}(z_2,\zb_2)\,\O^+_{\Delta_3}(z_3,\zb_3) \sim \frac{1}{z_{23}}\sum_{j=0}^\infty\frac{\Gamma(\Delta_2+j+1)\Gamma(\Delta_3-1)}{\Gamma(\Delta_2+\Delta_3+j)j!}\,\zb_{23}^j\bar{\partial}_3^j \O^-_{\Delta_2+\Delta_3-1}(z_3,\zb_3) + O(z_{23}^0)
\end{equation}
replicating the standard $+- \to -$ gluon OPE. Note that this matches the singular term in Equation \ref{defope}. 

\section{$S$-algebra}
\label{sec:S}

We now study the action of soft gluons inserted in leaf amplitudes on other gluon insertions. We define a soft gluon operator by 
\begin{equation}
    R^{k}(z,\zb) = \lim_{\Delta\to k}(\Delta-k)\O^ +_{\Delta}(z,\zb),
\end{equation}
which is a polynomial up to order $1-k$ in $\bz$ \cite{Guevara:2021abz}. 
By taking the soft limit of the OPEs in \eqref{eq:ppope} and \eqref{eq:pmope}, which contain all the $z-w$ poles, we see that 
\begin{equation}
    \begin{split}
        \oint_w\frac{\d z}{2\pi\im}\, R^k(z,\zb)\, \O^+_{\Delta}(w,\bw) &= \sum_{j=0}^{1-k} \frac{(-1)^{1-j-k}}{(1-j-k)!j!}\,\frac{\Gamma(\Delta-1)}{\Gamma(\Delta+j+k-2)}\,(\bz-\bw)^j\bar{\partial}^j\O^+_{\Delta+k-1}(w,\bw) \\
        \oint_w \frac{\d z}{2\pi\im}\, R^k(z,\zb)\,\O^-_\Delta(w,\bw) &= \sum_{j=0}^{1-k}\frac{(-1)^{1-j-k}}{(1-j-k)!j!}\,\frac{\Gamma(\Delta+1)}{\Gamma(\Delta+j+k)}\,(\zb-\bw)^j\bar{\partial}^j\O^-_{\Delta+k-1}(w,\bw)
    \end{split}
\end{equation}
forming the expected action of a soft gluon on a hard gluon.\footnote{We note that \eqref{defope} gives the primary terms in the OPE 
    $R^k(z_1,\zb_1)\,\O^-_{\Delta}(z_2,\zb_2) \sim \frac{1}{z_{12}}\left(\frac{(-1)^{1-k}\Gamma(\Delta+1)}{(1-k)!\Gamma(\Delta+k)}\O^-_{\Delta+k-1} +(-1)^{1-k} \sum_{n=1}^{1-k}\frac{\Gamma(\Delta+n+1)\Gamma(\Delta+n+k-1)}{(1-k-n)!n!\Gamma(\Delta+k+2n-1)\Gamma(\Delta+k+2n)}(z_{12}\zb_{12})^n\O^-_{\Delta+k+2n-1}\right)$. These $1-k$ extra primary terms, and their ++ counterparts we have not computed,  enter the soft theorems but do not have $z_{12}$ poles and do not deform the $S$-algebra. }

Additionally, by taking successive soft limits of the above expression, we can read off the standard soft gluon-soft gluon algebra \cite{Guevara:2021abz}
\begin{equation}\label{dsc}
    \oint_w \frac{\d z}{2\pi\im}\, R^k(z,\zb)R^\ell(w,\bw) = \sum_{j=0}^{1-k}\frac{(2-k-\ell-j)!}{(1-j-k)!(1-\ell)!}\frac{(\bz-\bw)^j}{j!}\,\bar{\partial}^j R^{k+j-1}(w,\bw)
\end{equation}
After certain rescalings of the operators, this can be identified with the loop algebra of Maps$(\C^2,\mathfrak{g})$, where $\mathfrak{g}$ is the Lie algebra of the gauge group, known as the `$S$-algebra' \cite{Strominger:2021mtt}. Hence the leaf amplitudes are governed by the same infinite-dimensional gauge symmetry algebra as the full celestial amplitudes, but not by spacetime translation invariance. 


\section*{Acknowledgements}

This work was supported by DOE grant de-sc/0007870, NSF GRFP grant DGE1745303, the Simons Collaboration on Celestial Holography and the Gordon and Betty Moore Foundation and the John Templeton Foundation via the Black Hole Initiative.

\bibliographystyle{JHEP}
\bibliography{refs}

\end{document}